\documentclass[twoside]{article}
\usepackage{fleqn,espcrc2}
\newcommand{\be}{\begin{equation}}
\newcommand{\ee}{\end{equation}}

\def\bea{\begin{eqnarray}}
\def\eea{\end{eqnarray}}
\def\thru#1{\mathrel{\mathop{#1\!\!\!\!/}}}
\def\pa#1{\frac{\partial}{\partial z^{}_#1}}
\def\pab#1{\frac{\partial}{\partial \overline z^{}_#1}}
\def\part#1{\partial^{}_#1}
\def\partb#1{\overline\partial^{}_#1}

\title{Boson-Fermion Confusion: The String Path To Supersymmetry}

\author{P. Ramond
\address{Institute for Fundamental Theory, \\
        Physics Department, University of Florida \\ 
       P.O. Box 118440, Gainesville, Fl 32605}
\thanks{Research supported in part by The United States Department 
of Energy grant DE-FG02-97ER41029.}}        
       
\begin{document}

\begin{abstract}
Reminiscences on the String origins of Supersymmetry are followed by a 
discussion of  the importance of confusing bosons with fermions in building 
superstring theories in $9+1$ dimensions.  In eleven dimensions, the kinship 
between bosons and fermions is more subtle, and may involve the exceptional group $F_4$.  
\vspace{1pc}
\end{abstract}

\maketitle

\section{Introduction}
Although the idea of a symmetry between bosons and fermions must be very old, 
after all both are present in Nature, I am only aware of non-relativistic 
attempts in that direction in the  1960's.  Stavraki~\cite{STA} (1967) 
proposed a current algebra with both commutators and anticommutators.  
Myazawa~\cite{MIYA} (1968), then at the University of Chicago, put 
together the fermions in the $\bf 56$ with the bosons in the $\bf 35$ of 
$SU(6)$ into one algebraic structure, which led him to invent the superalgebra 
we now call  $SU(6/21)$!   

In the Soviet Union,  a brilliant generalization of the Poincar\'e group  
to include anticommuting charges is proposed in 1971 by Golfand and Likhtman~\cite{GL},  
realizing relativistic supersymmetry in $3+1$ dimensions for the first time.  
This is followed in 1972 by a non-linear realization of this symmetry by Volkov and Akulov~\cite{VA}. 

String Theory stems from the  Dual Resonance Model, formulated to satisfy  
Dolen-Horn-Schmid duality~\cite{DHS}, according to which, in pion-nucleon scattering, 
the averaged s-channel fermionic resonances were  related to the t-channel boson 
exchanges. As such it implied a relation between bosons and fermions, although  
early workers seem to have put aside spin as an inessential complication. 
Amplitudes involving only bosons, known today as 
the Veneziano (open string) and Virasoro-Shapiro (closed string) models, 
are the progenitors of modern string theories. The    generalization  of  
Dual Models to include half-odd integer spins~\cite{R,NS} followed soon after in early 1971. 
  When expressed in terms of world-sheet symmetries, 
both R and NS formulations were shown~\cite{GS} to be examples of supersymmetry 
on the $1+1$ world-sheet. This in turn led to supersymmetry in $3+1$ 
dimensions, with explicit local interacting field-theories~\cite{WZ}. 
It is only 
later that space-time supersymmetry between the R and NS sectors was 
realized in $9+1$ dimensions by the GSO~\cite{GSO} projection. Born 
in the context of Dual Resonance Models, relativistic supersymmetry 
has now been found to play a central role in the formulation of 
quantum field theories, and perhaps even of Nature itself. 

\section{Reminiscences}
This conference affected me like a {\it madeleine}, and although still 
quite young,  I will take a few lines to offer some recollections 
of the epic period when the building blocks of supersymmetry were being 
laid down. Soon after  graduating from Syracuse University in spring  1969, 
my wife and I sailed to Europe to spend the summer at the ICTP, where J. Nuyts and H. 
Sugawara introduced me to 
the joys of working on the Veneziano model. By the time we sailed back 
to the United States to join the 
Fermi National Laboratory (then known as NAL) as one of its first 
postdocs, I had been hopelessly seduced by their  elegance and promise 
of simplicity. So much so that,  in the 
middle of the Atlantic, I found myself  in the reading room of the liner {\it France}, 
with a seminal paper by Fubini and Veneziano on the harmonic oscillator 
formulation. I then went to my cabin 
to fetch  postcards. Upon my return I saw the Fubini-Veneziano 
preprint on the table I had vacated.  Doubly puzzled,  since I had 
just left it in my cabin, and   it was, unlike my copy,  heavily 
annotated, I decided to wait for its owner. In walks a rather 
lanky individual to reclaim his posession; it was Andr\'e Neveu, on his 
way  to join Jo\"el Scherk in Princeton! 

In the Fall of 1969, at NAL, I started  working with Lou Clavelli on 
the group theoretical structure of the Veneziano amplitudes. There were no 
senior theorists at NAL; we were on our own, but for Nambu, who   
 was very supportive of our work, and even invited us to lunch at the Quadrangle Club! 
I also met  in that period, many of the early luminaries: M. Virasoro, 
then at Madison, Scherk and his first wife (whom he 
had met at the Club M\'editerran\'ee!),  S. Fubini, B. Sakita,  and G. Veneziano. 

In the spring of 1970, the Director of NAL, Bob Wilson decreed:  
``All theorists must go to Aspen". I had never heard of the place, 
did not want to go, but one did not resist Bob very long. That summer, at the 
Aspen Center for Physics, the break from grungy 
calculations, combined with the mountain air and the wonderful 
music, made me realize an easy way to look at 
these models. I decided to use the generalized position and 
momentum operators, introduced by Fubini and Veneziano,  in 
devising equations of motion. Upon my return at NAL, I wrote 
a short paper on the Dual Klein-Gordon equation, where I spoke 
of the correspondance between Dual systems and point particles 
(later to be known as the zero-slope limit). To my dismay, it 
was rejected by Physics Letters, and I withdrew the paper. Shaken 
but undaunted, I  proceeded to generalize the Dirac equation. 
There I found to my surprise a new algebraic structure, 
the square root of the Virasoro algebra. That Fall, I 
went to the Institute in Princeton where Nambu was spending 
a sabbatical. Again he was very supportive of my ideas, and 
encouraged me further, but I got stuck because I did not know 
Grassmann variables! 
It is during that visit that I met Mike Green, but I had to wait 
until the spring of 1971 to meet 
John Schwarz, and even longer L. Brink, and C. Thorn,  my lifelong dual friends!

\section{The First Fermion-Boson  Confusion}
Fermions and bosons cannot easily be confused, as they 
differ both by  their quantization and space-transformation properties, but 
the latter can be very similar in some dimensions. 
An  early example, where  assigning bosonic quantum numbers to fermions leads 
to a relativistic theory, is the introduction of fermions to Dual Models~\cite{R}. The 
$N$-point (bosonic) Veneziano amplitude is a sum of terms of the form

\be
<\,1\,\vert V(2)V(3)\cdots V(N-1)\vert\,N\,>\ ,
\ee
where 
\be
V(j)\sim~:e^{ik_j\cdot Q(\theta_j)}:\ ,
\ee
is the vertex for the emission of the $jth$ particle of momentum $k_j$, which 
 resembles a plane wave with a generalized coordinate 
\be
Q_\mu(\theta)=x_\mu+\theta p_\mu+\sum_1^\infty\frac{1}{\sqrt{n}}
\Big(a_\mu^{(n)\dagger}e^{-in\theta}+a_\mu^{(n)}e^{in\theta}\Big)\ .\ee
Here $x_\mu$ and $p_\mu$ are the usual position and momentum, 
and there are an infinite number of relativistic harmonic oscillators
\be
[\,a_\mu^{(n)}, a_\rho^{(m)\dagger}\,]=\delta^{nm}g_{\mu\rho}\ .\ee
This leads to a generalized momentum
\be
P_\mu(\theta)=\frac{dQ_\mu(\theta)}{d\theta}~=~
p_\mu-i\sum_1^\infty\sqrt{n}\Big(a_\mu^{(n)\dagger}e^{-in\theta}-a_\mu^{(n)}e^{in\theta}\Big)\ .\ee
Both  operators reduce to their point limit in the  (zero slope) limit
\be
Q_\mu(\theta)~~\rightarrow x_\mu\ ;\qquad P_\mu(\theta)~~\rightarrow p_\mu\ .\ee
Now if we apply this correspondance to the Klein-Gordon operator,  
\be
p_\mu\,p^\mu\,\Phi=0~~\rightarrow P_\mu(\theta)\,P^\mu(\theta)\,\Phi=0\ ,\ee
with
\be
P_\mu(\theta)\,P^\mu(\theta)=L_0+\sum_1^\infty\Big(L_ne^{-in\theta} +L_{-n}e^{in\theta}\Big)\ ,\ee
where the $L$'s satisfy the Virasoro algebra. In particular, $L_0$ is, up to an additive constant, 
the equation of motion for the bosons, and the $L_n\sim p\cdot a^{(n)}$ 
are akin to the decoupling operators of QED. Hence the analogy is 
established: take the generalized quantities, and replace the 
product of the averages by the average of the products:
\be
<\,P_\mu(\theta)\,>\,<\,P^\mu(\theta)\,>~~\rightarrow ~~
<\,P_\mu(\theta)\,P^\mu(\theta)\,>\ ,\ee
where $<\,\dots\,>$ denotes the average, or integration over $\theta$. 

The same procedure can be applied to the Dirac equation
\be
\gamma\cdot p\,\Psi~=~0\ ,\ee
by imagining that the gamma matrices are themselves averages of something
\be
\gamma_\mu~~\rightarrow ~~\Gamma_\mu(\theta)~=~\gamma_\mu+\cdots\ ,\ee
leading to a generalized Dirac equation 
\be
<\,\Gamma(\theta)\cdot P(\theta)\,>\Psi~=~0\ ,\ee
together with
\be
\{\,\Gamma_\mu(\theta),\Gamma_\rho(\theta')\,\}~=~2g_{\mu\rho}
\delta(\theta-\theta')\ .\ee
this requires the introduction of anticommuting harmonic oscillators 
which carry  {\it vector} indices
\be
\Gamma_\mu(\theta)=\gamma_\mu+\gamma_5\sum_1^\infty\Big(b_\mu^{(n)\dagger}
e^{-in\theta}+b_\mu^{(n)}e^{in\theta}\Big)\ ,\ee
 where $\gamma_5$ anticommutes with the Dirac matrices. 
Then  
\be
\Gamma(\theta)\cdot P(\theta)=F_0+\sum_1^\infty\Big(F_ne^{in\theta}+
F_{-n}e^{-in\theta}\Big)\ ,\ee
where the $F$'s satisfy the superVirasoro algebra. This chain of reasoning 
covers the genesis of the Dirac equation and the appearance of commuting 
and anticommuting structures in a relativistic framework. It is merely a 
generalization of the algebra of  Dirac's operator
\be
\{\,\gamma\cdot p\,,\,\gamma\cdot p\,\}~=~2\,p\cdot p\ .\ee
This procedure, assigning a vector index to a fermion, might appear foolish 
at first glance, except in $9+1$ dimensions where the light-cone little 
group is $SO(8)$, spinors and vectors have the same number of degrees of freedom. 

In relativistic theories,  bosons and fermions usually transform differently 
under space rotations. For massless particles the relevant group of 
rotations is the light-cone little group. In $1+1$ dimensions, there 
is no such group, and bosons  differ from fermions only by  
quantization, and  one can build bosons out of fermions 
without group-theoretical obstructions; the same applies to 
$2+1$ dimensions. In $3+1$ dimensions, the little group 
is non-trivial and fermions are distinguished by their 
helicities-- integer for bosons, half-integer for fermions. 
In higher dimensions,  fermions (bosons) transform according 
to the spinor (tensor)  representations of the Non-Abelian 
little group. In $9+1$ dimensions, the massless little group 
is $SO(8)$ and bosons and fermions have the same number of 
degrees of freedom. This fact lies 
at the heart of the superstring constructions. In $10+1$ 
dimensions and above, they become different again. However, 
in special numbers of dimensions, a strange kinship between 
spinor and tensor representations of the appropriate rotation 
group appears. In eleven dimensions, it leads to the 
supergravity theory, and, as we will show, possibly more.

\section{A Second Fermion-Boson  Confusion?}
In $10+1$ dimensions, there is no  apparent kinship between fermions and bosons. 
Yet there exists a supersymmetric theory in eleven dimensions, M-theory, 
with supergravity as its local limit. The degrees of freedom of supergravity 
are massless particles, belonging to  representations of $SO(9)$, the light-cone 
little group:
\begin{itemize}
\item Graviton as a symmetric second-rank tensor $h_{(ij)}$, with Dynkin label $[2000]$, 
\item Third-rank antisymmetric tensor,  $A_{[ijk]}$, with Dynkin label $[0010]$,
\item Rarita-Schwinger spinor-vector,  $\Psi_{\alpha i}$, with label $[1001]$.
\end{itemize} 
Their group-theoretical properties are summarized in the following table 
of Dynkin indices of different orders~\cite{PATERA}
\hskip 2cm
\begin{center}
\begin{tabular}{|c|c|c|c|}
\hline
$~{\rm irrep}~$& $[1001]$&$ [2000]$ & $[0010]$   \\
 \hline \hline         
$~I_0~ $&$  128$ & $ 44$ & $ 84$  \\
 \hline    
$~I_2~$& $256$& $88$ & $168$  \\
\hline
$~I_4~$& $640$& $232$ & $408$ \\                                                            
 \hline
$~I_6~$&$1792$& $712$ &$1080$\\ 
\hline
$~I_8~$&$5248$& $2440$ &$3000$\\ 
\hline
 \end{tabular}\end{center}
\vskip 0.3cm
These indices, except for $I_8$, match between the fermion and the two bosons. It turns 
out that there are infinitely many trios of representations of $SO(9)$ with similar 
group-theoretic relations among them. The simplest example is given by 
the triplet made of fields with index structure
\be h_{(ijk)l}+A_{(ij)(kl)m}+\Psi_{\alpha (ij)k}\ ,\ee 
and group-theoretical properties 
\hskip 2cm
\begin{center}
\begin{tabular}{|c|c|c|c|}
\hline
$~{\rm irrep}~$&$ [2100]$ & $[0110]$ & $[1101]$  \\
 \hline \hline         
$~I_0~ $&$  910$ & $1650$ & $ 2560$  \\
 \hline    
$~I_2~$& $3640$& $6600$ & $10240$  \\
\hline
$~I_4~$& $19864$& $34920$ & $54784$ \\                                                            
 \hline
$~I_6~$&$130840$& $217320$ &$348160$\\ 
\hline
$~I_8~$&$977944$& $1498344$ &$2466304$\\ 
\hline
 \end{tabular}\end{center}
\vskip 0.3cm 
They describe higher spin massless fields, with no apparent supersymmetry. 
This is only one example of this infinite set, which can be obtained from 
 a character formula\cite{GKRS},  traced to the three equivalent 
embeddings of $SO(9)$ inside the exceptional group $F_4$! Under the embedding 
$F_4\supset SO(9)$, the $52$ parameters of $F_4$ contain the $36$ generators 
of $SO(9)$ and $16$ parameters which transform as the $SO(9)$ spinor 
representation, and label the coset $F_4/SO_9$. Kostant\cite{KOS} 
introduces over that space sixteen $(256\times 256)$ gamma matrices 
which generate the Clifford algebra
\be
\{\,\gamma^a\,,\,\gamma^b\,\}~=~2\,\delta^{ab}\ ,~~~a,b=1,2\dots,16\ .
\ee
Note that the ``vector indices" of these matrices actually transform as 
the spinor of $SO(9)$! This is possible because of the anomalous embedding 
$SO(16)\supset SO(9)$, where the $16$ vector of $SO(16)$ is the $16$ 
spinor of $SO9)$. Another example of fermion-boson confusion. Let $T^a$ be 
the generators of $F_4$ not contained in $SO(9)$, and form the Kostant equation
\be
\sum_1^{16}\gamma^a\,T^a\,\Psi~\equiv~ \thru {\cal K}\,\Psi=~0\ .
\ee
Its solutions consist of all triples, including the supergravity multiplet. 
It is convenient to rewrite the gamma matrices in terms of eight Grassmann 
variables, and express the solutions as chiral superfields  in these variables,
\be
\Psi~=~\psi_0+\psi_i\theta_1+\psi_{ij}\theta_i\theta_j+\cdots \ ,\ee
and the supergravity solution corresponds to all  constant $\psi$. 
Under $SO(9)\supset SO(7)\times SO(2)$, these split as
\be
{\bf 1}+{\bf 8}+{\bf 28 }+{\bf 56}+{\bf 70}+{\bf 56}+{\bf 28}+\overline{\bf 8}+{\bf 1}\ ,\ee
reproducing the supergravity multiplet. Other solutions involve 
fields of higher spin. If these fields are to be incorporated in a 
relativistic theory, we must overcome the problem of massless spins 
with spins higher than two, bringing in well documented difficulties, 
with  coupling spin-one current~\cite{CG} and energy-momentum 
tensor~\cite{WW,AD} to massless particles of  spin greater than one. 

\section{A Simpler Example}
A similar but much simpler  construction can be achieved for the coset 
$SU(3)/SU(2)\times U(1)$. At the lowest level, it leads to a triplet of representations 
on which $N=2$ supersymmetry can be realized in $3+1$ dimensions.    

\subsection{The $N=2$ Hypermultiplet}
We first recall  the 
well-known light-cone description of the massless $N=2$ hypermultiplet in $3+1$ 
dimensions~\cite{BBB}, which  contains two Weyl spinors 
and two complex scalar fields, on which the $N=2$ SuperPoincar\'e algebra is realized. 
Introduce the  light-cone Hamiltonian
\be
P_{}^-=\frac{p\overline p}{2p^+}\ ,\ee
where $p=\frac{1}{\sqrt{2}}(p_{}^1+ip_{}^2)\nonumber \ .$ 
The front-form supersymmetry generators satisfy the anticommutation relations
\bea \nonumber \{{\cal Q}^{m}_+ ,\overline {\cal Q}^{n}_+\}&=&-2\delta_{}^{mn}p_{}^+\ ,\\
\{{\cal Q}^{m}_- ,\overline {\cal Q}^{n}_-\}&=&-2\delta_{}^{mn}\frac{p\overline p}{p_{}^+}\ ,~~~~~m,n=1,2\ ,\\
\nonumber \{{\cal Q}^{m}_+ ,\overline {\cal Q}^{n}_-\}&=&-2p\delta_{}^{mn}\ .\eea
The kinematic supersymmetries are expressed as 
\be
{\cal Q}^{m}_+=-\frac{\partial}{\partial\overline\theta^m}-\theta_m p_{}^+\ ,~~
\overline {\cal Q}^{m}_+=\frac{\partial}{\partial\theta^m}+\overline\theta_m p_{}^+\ ,\ee
while the kinematic Lorentz generators are given by
\bea\nonumber
M_{}^{12}&=&i(x\overline p-\overline xp)+\frac{1}{2}\theta_m
\frac{\partial}{\partial\theta_m}-\frac{1}{2}\overline\theta^m
\frac{\partial}{\partial\overline\theta^m}\ ,\\ \nonumber   
M_{}^{+-}&=&-x_{}^-p_{}^+ -\frac{i}{2}\theta_m\frac{\partial}
{\partial\theta_m}-\frac{i}{2}\overline\theta^m\frac{\partial}
{\partial\overline\theta^m}\ ,   \\ \nonumber
M_{}^{+}&\equiv&\frac{1}{\sqrt{2}}(M_{}^{+1}+iM_{}^{+2})=-xp_{}^+\ ,\\ 
\overline M_{}^{+}&=&-\overline xp_{}^+\ ,\\ \nonumber
\eea
where the  two complex Grassmann variables satisfy the anticommutation relations
\bea
\nonumber \{\theta_m,\frac{\partial}{\partial\theta_n} \}
&=&\{\overline\theta^m,\frac{\partial}{\partial\overline\theta^n}\}=\delta^{mn}\ ,\\ 
\nonumber \{\theta_m,\frac{\partial}{\partial\overline\theta^n} \}&=&
\{\overline\theta^m,\frac{\partial}{\partial\theta_n} \}
=0\ .\eea
The  (free) Hamiltonian-like supersymmetry generators are simply
\be
{\cal Q}^{m}_-=\frac{\overline p}{p_{}^+}{\cal Q}^{m}_+\ ,\qquad
\overline {\cal Q}^{m}_-=\frac{ p}{p_{}^+}\overline {\cal Q}^{m}_+\ ,\ee
and the light-cone boosts are given by
\bea
M_{}^{-}&=&x_{}^-p-\frac{1}{2}\{x,P_{}^-\}+i\frac{p}{p_{}^+}\theta_m
\frac{\partial}{\partial\theta_m}\ ,
\\ \nonumber
\overline M_{}^{-}&=&x_{}^-\overline p-\frac{1}{2}\{\overline x,P_{}^-\}+i\frac{
\overline p}{p_{}^+}\overline \theta^m
\frac{\partial}{\partial\overline \theta^m}\ ,\eea
where
$$ x=\frac{1}{\sqrt{2}}(x_{}^1+ix_{}^2)\ .$$
These generators represent the superPoincar\'e algebra on reducible superfields 
 because  the operators 
\be
{\cal D}_+^m~=\frac{\partial}{\partial\overline\theta^m}-\theta_m p_{}^+\ ,\ee
 anticommute with the supersymmetry generators. Irreducibility is achieved 
by acting on superfields for which 
\be
{\cal D}_+^m~\Phi=[\frac{\partial}{\partial\overline\theta^m}-\theta_m p_{}^+]\Phi =0\ ,\ee
solved by the chiral superfield

\be
\Phi(y^-,x^i,\theta_m)= \psi^{}_0+ \theta^{}_m \psi_{}^m + 
\theta^{}_1 \theta^{}_2\psi_{}^{12}\ ,\ee
where the arguments of the $\psi$'s depend on   
\be
y^-=x^--i\theta_m\overline\theta^m\ ,\ee
and the transverse variables.  Acting on this chiral superfield, the 
constraint is equivalent to requiring that

\be
{\cal Q}_+^m\approx -2p^+\theta_m\ ,\qquad \overline{\cal Q}_+^m\approx 
\frac{\partial}{\partial\theta_m}\ ,\ee
where the derivative is meant to act only on the naked $\theta_m$'s, 
not on those hiding in  $y^-$. 

\subsection{Coset Construction}
The degrees of freedom of the $N=2$ hypermultiplet in four dimensions appear 
as trivial solutions of the Kostant equation associated with the coset $SU(3)/SU(2)\times U(1)$. 
Let $T^A$ , $A=1,2,\dots, 8$, be the generators of $SU(3)$. Among
those, $T^i$, $i=1,2,3$,  and $T^8$   generate its $SU(2)\times U(1)$
subalgebra. Introduce  Dirac matrices over the coset 
\be \{\gamma^a, \gamma^b \} = 2\delta ^{ab}\ ,~~~a,b=4,5,6,7\ .\ee
 The Kostant equation over the coset $SU(3)/SU(2)\times U(1)$
\be
\thru {\cal K}\Psi~=~\sum_{a=4,5,6,7}\gamma_{}^aT^{}_a\Psi~=~0\ ,\nonumber \ee
has an infinite number of solutions which come in groups of three representations of 
$SU(2)\times U(1)$, called Euler triplets. For each representation of
$SU(3)$, there is a unique Euler triplet, $\{a_1,a_2\}$: 
\be[a^{}_2]_{-\frac{2a^{}_1+a_2+3}{6}}\oplus 
[a^{}_1+a_2+1]_{\frac{a^{}_1-a_2}{6}}\oplus [a^{}_1]_{\frac{2a^{}_2+a_1+3}{6}}\ ,\ee
where $a_1,a_2$ are the Dynkin labels of the associated $SU(3)$ representation. 
Here, $[a]$ stands for the $a=2j$ representation of $SU(2)$, and the
subscript denotes the $U(1)$ charge. Kostant's operator commutes with  
the $SU(2)\times U(1)$ generated by 
\be
L^{}_i=T^{}_i+S^{}_i\ ,~i=1,2,3\ ;\qquad L^{}_8=T^{}_8+S^{}_8\ ,\ee
a sum of the $SU(3)$ generators and the ``spin" part, expressed in terms 
of the $\gamma$-matrices as 
\be 
S^{}_j=- \frac{i}{4}f^{}_{jab}\gamma_{}^{ab}\ , \qquad S^{}_8=-
\frac{i}{4}f^{}_{8ab}\gamma_{}^{ab}\ ,\ee
where $\gamma^{ab}=\gamma^a\gamma^b\ ,a\ne b\ ,$ and $f^{}_{jab}\ ,
f^{}_{8ab}$ are structure functions of $SU(3)$. The Euler triplet 
corresponding to $a_1=a_2=0$,
\be\{0,0\}~=~[0]_{-\frac{1}{2}}\oplus [1]_{0}\oplus [0]_{\frac{1}{2}}\ ,\ee
describes the degrees of freedom of the $N=2$ supersymmetric multiplet,
when the $U(1)$ is interpreted as the helicity of the four-dimensional Poincar\'e algebra. 

Is it possible to  link this supersymmetric triplet to the others for 
which $a_{1,2}\ne 0$, while preserving  relativistic invariance? Not 
all triplets can decribe relativistic particles, since their $U(1)$ 
charges are in general fractional numbers, leading to states that 
pick up strange phases after a space rotation by $2\pi$, while Fermi-Dirac 
statistics only allows states for which this phase is $\pm 1$. Only Euler 
triplets for which  
\be a_1-a_2=3n\ ,\ee
where $n=0,\pm 1,\pm 2,\dots$, yield 
half-odd integer or integer $U(1)$ charges fit the bill. These Euler 
multiplets split into two groups, the self-conjugate, 
\be \{a,a\}~:~~~~[a]_{-\frac{a+1}{2}}\oplus [2a+1]_{0}\oplus [a]_{\frac{a+1}{2}}\ ,\ee
which contain equal number of half-odd integer-helicity fermions and
integer-helicity bosons, and naturally satisfy CPT.  The other possible Euler 
multiplets are of the form $\{a,a+3n\}$  with  $n=1,2,\dots$,
$$[a]_{-\frac{a+2n+1}{2}}\oplus [2a+3n+1]_{\frac{n}{2}}\oplus [a+3n]_{\frac{a+n+1}{2}}\ .$$ 
Since  CPT requires states of opposite helicity, these must be accompanied by their conjugates, 
 $\{a+3n,a\}$, with all helicities reversed. If both $n$ and $a$ are
even, each representation contains $(2a+3n+2)$ bosons and fermions, the
fermions appearing in two different $SU(2)$ representations. 

The helicities within each triplet are separated by more than
half a unit, and they cannot be related by  operations, such as supersymmetry,  
which change helicity by half a unit. 
Thus a necessary condition for supersymmetry
to be realized is to include all triplets, leading to an infinite-component theory.

We also note that there are states in the higher Euler triplets with  
helicities larger than $2$. If they are to be interpreted as massive 
relativistic states, they must arrange themselves in $SO(3)$ representations, 
which does not appear likely. Otherwise they  must
be interpreted as massless particles in four dimensions, lealing with a theory 
of massless states of spin higher than two. 

There are well-known difficulties with such theories\cite{WW,AD}. In particular, they do not 
have covariant  energy momentum tensors, and it must be that in the flat 
space limit they decouple from the gravitational sector. Alternatively, 
the no-go theorems do not apply if  there are an infinite number of 
such particles. The best argument against such theories is that no  
working example has yet been produced, but we hope  such a theory 
can be formulated with an infinite number of Euler multiplets. 

\subsection{Grassmann Numbers and  Dirac Matrices}
In order to make contact with the supersymmetry of the lowest Euler
triplet, we represent~\cite{BRINK} the $\gamma$-matrices in terms 
of Grassmann numbers and their derivatives as
\bea\nonumber
\gamma_{4+i5}&=&i\sqrt{\frac{2}{p^+}}{\cal Q}_+^1\ ,~~
\gamma_{4-i5}=i\sqrt{\frac{2}{p^+}}\overline{\cal Q}_+^1\\
\nonumber\gamma_{6+i7}&=&i\sqrt{\frac{2}{p^+}}{\cal Q}_+^2\ ,~~
\gamma_{6-i7}=i\sqrt{\frac{2}{p^+}}\overline{\cal Q}_+^2\ ,
\eea
in terms of the kinematic $N=2$ light-cone supersymmetry generators defined 
in the previous section. It follows that the Kostant operator anticommutes 
with the constraint operators
\be\{~\thru {\cal K},~{\cal D}_+^m\}~=~0\ ,\ee
so that we can simplify its solutions to chiral superfields, on which these become   
 \bea\nonumber
\gamma_{4+i5}&=&-2i\sqrt{2p^+}~\theta_1\ ,~~
\gamma_{4-i5}=i\sqrt{\frac{2}{p^+}}~\frac{\partial}{\partial\theta_1}\\
\nonumber\gamma_{6+i7}&=&-2i\sqrt{2p^+}~\theta_2\ ,~~
\gamma_{6-i7}=i\sqrt{\frac{2}{p^+}}~\frac{\partial}{\partial\theta_2}\ ,
\eea
The ``spin" parts of the $SU(2) \times U(1)$ generators,  expressed
in terms of  Grassmann variables, do not depend on $p^+$,  
\bea\nonumber
S^{}_1&=&\frac{1}{2}  ( \theta^{}_1\frac{\partial}{\partial
\theta^{}_2}+ 
\theta^{}_2\frac{\partial}{\partial \theta^{}_1})\ ,\\\nonumber
S^{}_2&=&-\frac{i}{2}  (\theta^{}_1 \frac{\partial}{\partial
\theta^{}_2}-  
\theta^{}_2 \frac{\partial}{\partial \theta^{}_1})\ ,\\\nonumber
S^{}_3&=&\frac{1}{2}  (\theta^{}_1\frac{\partial}{\partial
\theta^{}_1}-  
\theta^{}_2\frac{\partial}{\partial \theta^{}_2})\ ,\eea
and 
$$
S^{}_8=\frac{\sqrt3}{2} (\theta^{}_1 \frac{\partial}{\partial
\theta^{}_1}+
\theta^{}_2 \frac{\partial}{\partial \theta^{}_2}-1)\ ,
$$
identified with the helicity, up to a factor of $\sqrt{3}$. 

\subsection{Linear Realization of $SU(3)$}
The $SU(3)$ generators can be conveniently expressed on three complex
variables and their conjugates. Define for convenience the differential operators
$$\part1\equiv\pa1\ ,\qquad \partb1\equiv\pab1\ ,~{\rm etc.}\ ,$$
in terms of which the generators are given by 
$$ T^{}_1+iT^{}_2=z^{}_1\part2-\overline z^{}_2\partb1\ ,\qquad 
T^{}_1-iT^{}_2=z^{}_2\part1-\overline z^{}_1\partb2\ ,$$
$$T^{}_4+iT^{}_5=z^{}_1\part3-\overline z^{}_3\partb1\ ,\qquad 
T^{}_4-iT^{}_5=z^{}_3\part1-\overline z^{}_1\partb3\ ,$$
$$ T^{}_6+iT^{}_7=z^{}_2\part3-\overline z^{}_3\partb2\ ,\qquad 
T^{}_6-iT^{}_7=z^{}_3\part2-\overline z^{}_2\partb3\ ,$$
and
$$ T^{}_3=\frac{1}{2}(
z^{}_1\part1-z^{}_2\part2-\overline z^{}_1\partb1+\overline z^{}_2\partb2)\ ,$$
$$ T^{}_8=\frac{1}{2\sqrt{3}}(
z^{}_1\part1+z^{}_2\part2-\overline z^{}_1\partb1-\overline z^{}_2\partb2-2
z^{}_3\part3+2\overline z^{}_3\partb3)\ .$$
These  act as hermitian operators on holomorphic functions of
$z_{1,2,3}^{}$ 
and $\overline z_{1,2,3}^{}$, normalized with respect to the inner product

$$ (f,g)\equiv~\int d^3zd^3\overline z~e^{-\sum_i\vert z_i\vert^2}~
f^*(z,\overline z)~g(z,\overline z)\ .$$
It is convenient to introduce the positive integer Dynkin labels $a_1$ and $a_2$, for which 
$$ T^{}_3\vert~ a_1,a_2>=~\frac{a_1}{2}\vert~ a_1,a_2>\ ,$$
and
$$T^{}_8\vert~ a_1,a_2>~=~\frac{1}{2\sqrt{3}}(a_1+2a_2)\vert~ a_1,a_2>\ .$$
The highest-weight states of each $SU(3)$ representation are holomorphic 
polynomials of the form
$$ z^{a_1}_1\overline z_3^{a_2}\ ,$$
where $a_1,a_2$ are the Dynkin indices, since it is easily seen to reproduce 
the above values for $T_3$ and $T_8$. This  describes all
representations 
of $SU(3)$ as homogeneous holomorphic polynomials. Finally we note that any function of  the quadratic invariant

$$ Z^2\equiv \vert z_1^{}\vert^2+\vert z_2^{}\vert^2
+\vert z_3^{}\vert^2\ ,$$
 can  multiply these polynomials without affecting their $SU(3)$ transformation properties.

\subsection{Solutions of Kostant's Equation}
 Kostant's equation
$$\thru {\cal K}\Psi~=~\sum_{a=4,5,6,7}\gamma_{}^aT^{}_a\Psi~=~0\ .$$
now becomes two coupled systems of equations
\bea\nonumber (z^{}_1\part3-\overline z^{}_3\partb1)\psi^{}_1+(z^{}_2\part3-
\overline z^{}_3\partb2)\psi^{}_2&=&0\ ,\\ \nonumber(z^{}_3\part1-
\overline z^{}_1\partb3)\psi^{}_2-(z^{}_3\part2-\overline z^{}_2
\partb3)\psi^{}_1&=&0\ ,\eea
and
\bea\nonumber (z^{}_3\part1-\overline z^{}_1\partb3)\psi^{}_0-(z^{}_2\part3-
\overline z^{}_3\partb2)\psi^{}_{12}&=&0\ ,\\ \nonumber (z^{}_3\part2-
\overline z^{}_2\partb3)\psi^{}_0+(z^{}_1\part3-\overline z^{}_3\partb1)\psi^{}_{12}&=&0\ .\eea
The homogeneity operators
$$D=z^{}_1\part1+z^{}_2\part2+z^{}_3\part3\ ,\qquad \overline 
D=\overline z^{}_1\partb1+\overline z^{}_2\partb2+ \overline z^{}_3\partb3\,$$
commute with $\thru {\cal K}$, allowing the solutions of the Kostant
equation to  be arranged 
as irreps of the $SU(2)\times U(1)$ generated by the operators
$$L^{}_i=T^{}_i+S^{}_i\ ,~i=1,2,3\ ;\qquad L^{}_8=T^{}_8+S^{}_8\ .$$
The solutions for each triplet, conveniently written only for the highest weight states, are of the form 
\bea 
\Psi&=& z_3^{a_1}~\overline z_2^{a_2}\ :~~~~~[a_2]_{-\frac{2a^{}_1+a_2+3}{6}}
\ ,\cr
\Psi&=& \theta_1~z_1^{a_1}~\overline z_2^{a_2}\ :~~~~~
[a_1+a_2+1]_{\frac{a^{}_1-a_2}{6}}\ ,\cr
\Psi&=& \theta_1\theta_2~z_1^{a_1}~\overline z_3^{a_2}\ :~~~~
~[a_1]_{\frac{2a^{}_2+a_1+3}{6}}\ ,\eea
where we have indicated their $SU(2)$ Dynkin labels. All other
solutions in the same Euler triplet can be obtained by repeated 
action of the lowering operator 
$$L_1-iL_2= \theta^{}_2\frac{\partial}{\partial \theta^{}_1}+ 
(z^{}_2\part1-\overline z^{}_1\partb2)\ .$$
We now see how the triplets arise as polynomials of the same degree. 
\subsection{The Poincar\'e Algebra}
It is easy te represent the Poincar\'e algebra on the Euler triplets. Starting from 
the more general representation 
\bea\nonumber
M_{}^{12}&=&i(x\overline p-\overline xp)+\frac{1}{2}
\theta_m\frac{\partial}{\partial\theta_m}-\frac{1}{2}
\overline\theta^m\frac{\partial}{\partial\overline\theta^m}+S_{}^{12}\ ,\\ \nonumber   
M_{}^{+-}&=&-x_{}^-p_{}^+ -\frac{i}{2}\theta_m
\frac{\partial}{\partial\theta_m}-\frac{i}{2}
\overline\theta^m\frac{\partial}{\partial\overline\theta^m}\ ,   
\\ \nonumber
M_{}^{+}&=&-xp_{}^+\ ,\qquad 
\overline M_{}^{+}=-\overline xp_{}^+\ ,\\ \nonumber
M_{}^{-}&=&x_{}^-p-\frac{1}{2}\{x,P_{}^-\}+i\frac{p}{p_{}^+}(\theta_m
\frac{\partial}{\partial\theta_m}+S_{}^{12})\ ,
\\ \nonumber
\overline M_{}^{-}&=&x_{}^-\overline p-\frac{1}{2}\{\overline x,P_{}^-\}+i\frac{
\overline p}{p_{}^+}(\overline \theta^m
\frac{\partial}{\partial\overline \theta^m}-S_{}^{12})\ ,\eea
and identify the rest of the helicity generator  as
\be
\nonumber S_{}^{12}~=~\frac{1}{\sqrt{3}}T^{}_8\ .\ee
These act on chiral superfields whose entries are polynomials in $z$ and $\overline z$. 

There is no difficulty in writing a {\it free} Lagrangian for these solutions, 
but  one could contemplate writing a free Lagrangian for all the solutions at 
one fell swoop. For this, we need to consider a chiral superfield whose entries 
are arbitrary polynomials in the complex variables, but subject to the Kostant 
equation {\it as a constraint}. 
The action would be of the form
\be
S=\int d^2xdx^-\int d^3zd^3\overline z~{\cal L}\ ,\ee
suggesting that the classical space includes the group manifold as
well. The Lagrange density would then be built out of chiral
superfields of the form $\Psi(y^-,x^i,z_a,\overline z_a,\theta_i)$, 
which satisfy {\it both} the chiral and Kostant constraints. Work is in progress to see 
 if one can introduce interactions among these light-cone superfields .  
\section{Outlook}
This example only serves to introduce the method we intend to pursue. 
The interesting case in eleven dimensions 
singles out the coset $F_4/SO(9)$. A similar procedure will lead us to 
consider light-cone superfields over $4$ copies of $26$ real variables\cite{FUL}, 
and eight Grassmann variables. The light-cone Lorentz group generators 
are now built out of the $SO(9)$ generators, of the form
 \be
L^{ij}~=~T^{ij}+S^{ij}\ ,~~~i,j=1,2\dots, 9\ ,\ee
in which $T^{ij}$ generate the $SO(9)$ little group in the particular 
representation associated with  each Euler triplet, and they act along the $256\times 256$ unit matrix. In particular, $T^{ij}=0$ for the supergravity multiplet, and the ``spin" part is 
\be
S^{ij}~=~-\frac{i}{4}\sum_{a,b=1}^{16}\,f^{ij\,ab}\,\gamma^{ab}\ .\ee
Here the $f^{ij\,ab}$ are the structure functions of the exceptional group $F_4$!  In closing we note that the same coset plays a prominent role in the projective geometry associated with the 
Exceptional Jordan Algebra, leading one to hope in a new interpretation 
with $SO(9)$ as a space group.  Details will be presented elsewhere.

I wish to think Professors Shifman and Vainshtein for their warm and wonderful hospitality, 
and the opportunity to meet some pioneers of supersymmetry, and  re-acquaint myself with 
many  old friends. 
\vfill\eject

\end{document}